\pdfoutput=1
\documentclass[reprint,aps,prl,twocolumn,superscriptaddress]{revtex4-2}
\usepackage{graphicx}
\usepackage{dcolumn}
\usepackage{bm}
\usepackage{amsmath}
\usepackage[english]{babel}
\usepackage[utf8]{inputenc}
\usepackage{color}
\begin{document}

\title{Semiconductor materials stacks for quantum dot spin qubits}

\author{G. Scappucci}
\email{g.scappucci@tudelft.nl}
\affiliation{QuTech and Kavli Institute of Nanoscience, Delft University of Technology, PO Box 5046, 2600 GA Delft, The Netherlands}

\date{\today}

\begin{abstract}
In this perspective piece, I benchmark gallium arsenide, silicon, and germanium as material platforms for gate-defined quantum dot spin qubits. I focus on materials stacks, quantum dot architectures, bandstructure properties and qualifiers for disorder from electrical transport. This brief note is far from being exhaustive and should be considered a first introduction to the materials challenges and opportunities towards a larger spin qubit quantum processor.
\end{abstract}

\maketitle
\section{I\lowercase{ntroduction}}
A quantum computer capable of solving useful problems will require millions of excellent qubits\cite{campbell_roads_2017}. Spin qubits in gate-defined semiconductor quantum dots\cite{Loss1998} are a promising platform for large-scale integration of quantum components, due to the striking resemblance of quantum dot devices to the classical transistors\cite{vandersypen_interfacing_2017,vandersypen_quantum_2019} which are integrated by billions into a single chip. To host spin qubits in gate-defined quantum dots, a single electron or hole is trapped in the potential landscape obtained by a combination of heterogeneous materials, interfaces, and gate electrodes\cite{Hanson2007,Zwanenburg2013}. Here, I focus on GaAs, Si, and Ge, three semiconductors platforms that demonstrated the key functionality of two-qubit quantum logic\cite{nowack_single-shot_2011,veldhorst_two-qubit_2015,hendrickx_fast_2020}. 

\begin{figure*}[ht]
	\includegraphics[width=160mm]{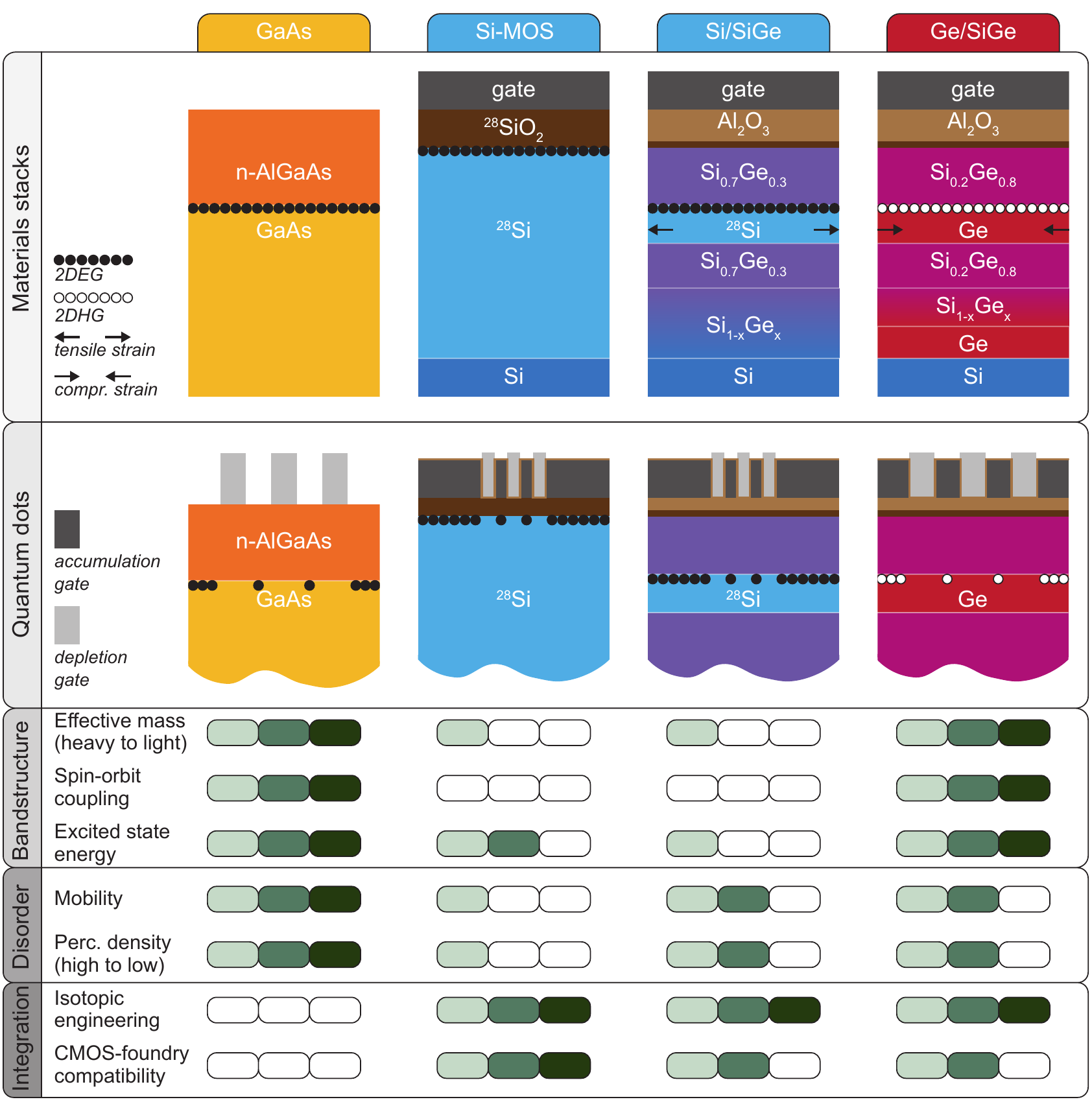}
	\caption{GaAs, Si, and Ge semiconductor materials for quantum dot spin qubits. The first and second row compare materials stacks and quantum dot architectures for GaAs, Si-MOS, Si/SiGe, and Ge/SiGe (left to right columns). In GaAs, the quantum well is typically 50-80~nm below the surface. In Si/SiGe and Ge/SiGe the distance from the quantum well to the dielectric interface is typically 30-60~nm. Ge quantum wells are thicker ($\approx$16~nm) than Si quantum wells ($\approx$10~nm). In quantum dots, the accumulation gate pitch is tighter in Si ($\approx$80~nm) than in Ge\cite{lawrie_quantum_2020}. A scorecard (three bottom rows) compares of useful properties that qualify material platforms for better quantum dot spin qubits. These properties arise from bandstructure (effective mass, spin-orbit coupling, excited state energy related to valley splitting), characterise disorder (mobility and percolation density), and are relevant for integration (isotopic engineering, CMOS-foundry compatibility). Full-score indicates that a property is beneficial for quantum-dot spin qubits. Spin-orbit coupling requires special consideration, since a trade-off is arguably desired between fast all-electrical local qubit driving and induced decoherence.}
\label{fig:MAT}
\end{figure*}

In Figure~\ref{fig:MAT}, the GaAs, Si, and Ge are qualitatively compared by considering materials stacks, quantum dot architectures, bandstructure properties, integration aspects and two electrical transport metrics---mobility and percolation density---that serve as useful qualifiers for disorder due to materials and gate stacks. Maximum mobility ($\mu$) is a popular metric to gauge the disorder potential landscape in the material. However, mobility peaks at high carrier density due to screening---not the regime of quantum dot qubits operation. The percolation density ($n_p$)\cite{Tracy2009ObservationMOSFET} is a better metric to characterize disorder relevant for quantum dots, since it measures the minimum density required to establish metallic conduction by overcoming charge trapping in the disorder potential landscape. Quantum mobility\cite{DasSarma2014MobilityStructures} and charge noise\cite{vandersypen_interfacing_2017} are two other key metrics that I intentionally omit here, because they require an in-depth and specialist discussion that goes beyond the scope of this brief perspective.

\section{G\lowercase{allium arsenide}}
Spin qubits in gate-defined quantum dots were first demonstrated in GaAs\cite{Petta2005,Hanson2007} due to favourable bandstructure properties and the maturity of molecular beam epitaxy of III-V compounds heterostructure. In a typical modulation-doped GaAs/AlGaAs heterostructure the conduction band offset at the GaAs/AlGas interface supports a two-dimensional electron gas (2DEG), populated from a nearby Si-doped AlGaAs layer. GaAs and AlGaAs are nearly lattice-matched, so GaAs/AlGas heterostructures have an exceptional structural quality, allowing for very large electron mobility (10$^6$--10$^7$~cm$^2$/Vs) and low percolation density ($<10^{10}$~cm$^{-2}$)\cite{manfra_transport_2007}. Consequently, quantum dots of about $1/\sqrt{n_p}=100$~nm in size, informative about the average distance between traps, are essentially disorder-free.  

To define quantum dots, the 2DEG is locally depleted by Schottky gates. The absence of dielectrics preserves the low disorder of the pristine heterostructure and increases device yield. The single conduction band valley and low effective mass ($m^*=0.067m_e$) enable quantum dots with large energy spacing thus easier to fabricate. The sizable spin-orbit interaction allows for local all-electrical manipulation of single spins\cite{Nowack2007}. The main drawback of GaAs is the hyperfine coupling to the nuclear spin bath, causing severe qubit decoherence. Furthermore, the challenging integration of GaAs on a Si wafer limits the prospect of integrating the very large number of qubits into a practical quantum processor.

\section{S\lowercase{ilicon}}
The two main limitations of GaAs become the main drivers to pursue spin qubits in Si. Isotopic enrichment into $^{28}$Si\cite{itoh_isotope_2014} drastically reduces hyperfine coupling providing long quantum coherence\cite{Loss1998,Zwanenburg2013,vandersypen_quantum_2019}. Furthermore, there is the promise to leverage advanced semiconductor manufacturing for the integration of qubits in the large numbers required for fault-tolerant quantum computing. Spin qubits in Si are implemented in metal oxide semiconductor (Si-MOS) structures or in Si/SiGe heterostructures.  In Si-MOS, a 2DEG is confined at the interface between an intrinsic $^{28}$Si epilayer and $^{28}$SiO$_2$. Multiple layers of gates, insulated by dielectrics, accumulate and confine charge into quantum dots\cite{angus_gate-defined_2007}. These gates have a rather tight pitch because of the large effective mass ($0.19m_e$), requiring quantum dots in Si to be much smaller than in GaAs. Following the availability of $^{28}$SiH$_4$ gas for $^{28}$Si deposition\cite{sabbagh_quantum_2019,mazzocchi_99.992_2019}, Si-MOS spin qubits were fabricated in a 300~mm semiconductor manufacturing facility using all-optical lithography and fully industrial processing\cite{zwerver2021qubits}. In these qubits, Si finFETs  provide conducting channels and gates on top induce quantum dots along the fin.

The presence of multiple conduction bands is a limitation of Si\cite{Zwanenburg2013}, because the small energy separation (valley splitting) between the ground state and the lowest excited state complicates quantum operations. However, due to the strong confinement at the sharp semiconductor/oxide interface, valley splitting in Si-MOS can be substantial\cite{yang_spin-valley_2013} (up to 1~meV), making it possible to operate Si qubits at "hot" temperatures of 1~K\cite{petit_universal_2020}. This is encouraging towards co-integration of CMOS-based cryogenic control circuits and silicon quantum processors\cite{xue2020cmos}. 

The main drawback of Si-MOS is the proximity of the qubit to the very disordered oxide interface. Mobility is rather low ($\approx 10^4$~cm$^2$/Vs at best\cite{shankar_lyon2010spin,Rochette2019,sabbagh_quantum_2019}) and the percolation density high ($\approx 10^{11}$~cm$^{-2}$\cite{sabbagh_quantum_2019}). Consequently, multi-quantum dot systems have proven challenging to control, at least for devices fabricated in academic settings.

Disorder is greatly mitigated in Si/SiGe heterostructures, because the 2DEG is accumulated at the buried interface between a tensile strained Si quantum well and a Si$_{1-x}$Ge$_x$ barrier (Ge concentration $x \approx 0.3$)\cite{schaffler_high-mobility_1997}. Like Si-MOS, the Si quantum well can be isotopically enriched to $^{28}$Si for long quantum coherence\cite{yoneda_quantum-dot_2018}. Differently than GaAs, Si/SiGe heterostructures for quantum dot spin qubits are undoped\cite{Maune2012CoherentDot} and electrons populate the quantum well via top-gates. High quality Si/SiGe poses additional challenges compared to GaAs/AlGaAs because of the 4.2\% lattice mismatch between Si and Ge. The Si quantum well is deposited on a strain-relaxed SiGe buffer obtained by gradually increasing the Ge concentration in the SiGe alloy to accommodate the lattice mismatch between Si and Ge. After decades of advancements, Si/SiGe heterostructures, grown by industrial reduced pressure chemical vapor deposition (RP-CVD), are a rather mature platform with high mobility ($\approx 10^5$~cm$^2$/Vs) and low percolation density ($\approx 10^10$~cm$^{-2}$)\cite{Mi2015InvestigationHeterostructures,Mi2017CircuitSilicon,wuetz2019multiplexed}), expected to further improve by leveraging advanced semiconductor manufacturing processes for optimising the gate stack. Fabrication of quantum dots in Si/SiGe relies on overlapping gate structures for tightly spaced quantum dots with gate tunable tunnel barriers. Due to the low disorder, device yield is high, making it possible to define and control large linear arrays of quantum dots\cite{zajac_scalable_2016}.

The current major challenge with Si/SiGe is the lower valley splitting (50-200~$\mu$eV\cite{watson_programmable_2018,borselli_pauli_2011,hollmann2020large,scarlino2017dressed}) compared to Si-MOS, due to the atomistic imperfection and chemical disorder at the epitaxial Si/SiGe interface. However, first steps are being taken to innovate the heterostructures by incorporating more complex Ge concentration profiles\cite{Neyens2018TheWells} for increased the valley splitting. 

\section{G\lowercase{ermanium}}
Whilst most studies have focused on electrons, holes in strained Ge/SiGe heterostructures have recently emerged as a compelling platform that offers low disorder, all-electrical qubit control, and avenue for scaling\cite{scappucci_germanium_2020}. Ge combines many advantages of Si and GaAs whilst overcoming most of their limitations. Ge is a CMOS-foundry material and can be isotopically engineered for long quantum coherence. The high mobility, low effective mass, and sizable spin-orbit coupling imply large energy spacing for easy fabrication of quantum dots with full electrical qubit control. 

In a typical Ge/SiGe heterostructure\cite{sammak_shallow_2019} holes are confined at the interface between a compressively-strained Ge quantum well and a Si$_{1-x}$Ge$_{x}$ barrier ($x\approx 0.8$). Strain and size quantization remove the valence band degeneracy, so holes in Ge/SiGe are a single-band system, overcoming the main limitation of electrons in Si. Starting from a Si wafer, a strain-relaxed Ge layer is grown, followed by a graded Si$_{1-x}$Ge$_{x}$ layer in which the Ge concentration is progressively reduced to $\approx 0.8$ (reverse grading). The strain-relaxed SiGe buffer serves as a virtual substrate for the coherent deposition of the Ge quantum well, the SiGe barrier, and a final sacrificial Si cap. Modulation doping is avoided and charge carriers populate the quantum well via top-gates. Optimized Ge/SiGe stacks are grown by RP-CVD and have similar levels of disorder as Si/SiGe, with $\mu \approx 10^5$~cm$^2$/Vs\cite{sammak_shallow_2019}) and $n_p\approx 10^{10}$~cm$^{-2}$\cite{lodari_low_2020}.

Quantum dot qubits\cite{hendrickx_fast_2020} are defined by a set of multi-layer gates, with less-stringent pitch than in Si, due to the light effective mass ($0.05m_e$\cite{lodari_light_2019}). The low disorder in Ge/SiGe has allowed for the rapid progress in only two years from single quantum dots\cite{hendrickx_gate-controlled_2018} to a four qubit system\cite{hendrickx_four-qubit_2020} with controllable coupling along both directions of a 2$\times$2 array, setting the benchmark for spin qubit quantum processors. 

\section{O\lowercase{utlook}}
More than fifteen years after the first demonstration of spin qubits, it is unclear which material will be powering a large scale spin qubit quantum processor. However, the most significant advancements in the field can be traced back to leaps in materials developments. Silicon took over GaAs because the integration of $^{28}$Si enabled long quantum coherence. Germanium is now in the spotlight since hole spin qubits have progressed at an extraordinary pace. The Ge qubit count has doubled roughly every year and larger systems are on the horizon. The Ge quantum information route\cite{scappucci_germanium_2020} is poised to retain many advantages of GaAs and Si, whilst overcoming some of their respective long-standing challenges. 

Regardless of the material of choice, increasingly fast feedback cycles are required to accelerate the development of quantum materials. Important steps in this direction have been taken. Cryo-multiplexing technology \cite{wuetz2019multiplexed,pauka_cryogenic_2019} mitigates the interconnect bottleneck present in cryostats operating at mK. Therefore, we can access with high-throughput the low-temperature quantum transport properties of 2D electron or holes relevant for spin qubits. Hopefully, high-throughput characterisation will also apply to  charge noise measurements. Fast optimization of the material and gate stack parameters is essential as we are moving into the next phase of engineering qubit systems in the large numbers required for useful quantum computing.

\bibliography{bibliography.bib}

\end{document}